\documentclass[lettersize,journal]{IEEEtran}
\usepackage{amsmath,amsfonts}
\usepackage{array}
\usepackage{textcomp}
\usepackage{stfloats}
\usepackage{url}
\usepackage{verbatim}
\usepackage{graphicx}
\def\BibTeX{{\rm B\kern-.05em{\sc i\kern-.025em b}\kern-.08em
    T\kern-.1667em\lower.7ex\hbox{E}\kern-.125emX}}
\usepackage{balance}
\usepackage{graphicx,cite,calc,dsfont}
\usepackage{hyperref}
\usepackage{amsmath}
\usepackage{amsthm,amssymb}
\usepackage{algorithm}
\usepackage{algpseudocode}

\theoremstyle{definition}

\usepackage{empheq}
\usepackage{breqn}
\usepackage{comment}
\usepackage{threeparttable}
\usepackage{subfigure}

\include{macros}

\newcommand{\bit}{\begin{itemize}}
\newcommand{\eit}{\end{itemize}}

\newcommand{\bc}{\begin{center}}
\newcommand{\ec}{\end{center}}

\newcommand{\ba}{\begin{array}}
\newcommand{\ea}{\end{array}}

\newcommand{\beq}{\begin{equation}}
\newcommand{\eeq}{\end{equation}}

\newcommand{\beqn}{\begin{equation*}}
\newcommand{\eeqn}{\end{equation*}}

\newcommand{\bean}{\begin{eqnarray*}}
\newcommand{\eean}{\end{eqnarray*}}
\newcommand{\bea}{\begin{eqnarray}}
\newcommand{\eea}{\end{eqnarray}}

\begin{document}
\title{Adaptive Multi-Armed Bandit Learning for Task Offloading in Edge Computing}
\author{Lin Wang, Jingjing Zhang
\thanks{Lin Wang and Jingjing Zhang are with the Department of Communication Science and Engineering, Fudan University, Shanghai 200433, China (e-mail: 22210720073@m.fudan.edu.cn; jingjingzhang@fudan.edu.cn). This work has been supported by the National Natural Science Foundation of China Grant No. 62101134. \emph{(Corresponding author: Jingjing Zhang)}}
}

\maketitle

\begin{abstract}
The widespread adoption of edge computing has emerged as a prominent trend for alleviating task processing delays and reducing energy consumption. However, the dynamic nature of network conditions and the varying computation capacities of edge servers (ESs) can introduce disparities between computation loads and available computing resources in edge computing networks, potentially leading to inadequate service quality. To address this challenge, this paper investigates a practical scenario characterized by dynamic task offloading. Initially, we examine traditional Multi-armed Bandit (MAB) algorithms, namely the $\varepsilon$-greedy algorithm and the UCB1-based algorithm. However, both algorithms exhibit certain weaknesses in effectively addressing the tidal data traffic patterns. Consequently, based on MAB, we propose an adaptive task offloading algorithm (ATOA) that overcomes these limitations. By conducting extensive simulations, we demonstrate the superiority of our ATOA solution in reducing task processing latency compared to conventional MAB methods. This substantiates the effectiveness of our approach in enhancing the performance of edge computing networks and improving overall service quality.
\end{abstract}



\section{Introduction}
\IEEEPARstart{W}{ith} the rapid development of fifth-generation (5G) communication technology and the increasing prevalence of smart devices, the demand for data processing has grown exponentially. In this context, mobile edge computing (MEC) plays a crucial role in supporting time-sensitive applications and enhancing the quality of services (QoS) \cite{ref1, ref2}. By offloading tasks from end users, encompassing various intelligent devices, to edge servers (ESs) or the cloud, MEC effectively improves network efficiency and reduces response delays.

However, in the MEC scenario, ESs often exhibit variations in computation capacity, leading to potential mismatches between computation load and available computing resources. This discrepancy can result in reduced resource utilization efficiency and increased task processing latency. Consequently, it is essential to design effective task offloading and resource allocation strategies in MEC \cite{ref3}.


Numerous methods have been developed in prior research to optimize task offloading strategies, with a primary focus on reducing task processing latency and energy consumption within the MEC scenario, involving single or multiple ESs \cite{ref9, ref11, ref13, ref14, mao2016dynamic}. When an ES serves only one user, the central concern is to find a wise task offloading strategy for each end user that reduces delay or energy consumption. 
However, when an ES can serve multiple users simultaneously, there can be disparities between its computation loads and available computing resources, hence the necessity to design a reasonable joint task offloading policy. In this context, collaborative cache allocation and computation offloading schemes were proposed in Reference \cite{ref14}, where ESs collaborate in executing computation tasks and caching data. Additionally, reinforcement learning algorithms, which have gained substantial traction in recent times, have been increasingly employed in IoT networks to address resource allocation problems\cite{ref7}.


The MAB problem has received extensive attention in addressing the trade-off between exploration and exploitation in decision-making processes, especially in the presence of environmental uncertainty.  In the context of edge computing networks, MAB learning has proven to be valuable for identifying optimal task offloading and resource allocation strategies \cite{ref9, ref11, ref19, ref23}. For example, Miao \cite{ref9} developed an intelligent task caching algorithm that utilizes MAB methods to dynamically adjust the caching strategy based on task size and computing requirements, resulting in a significant reduction in average task latency. In \cite{ref19}, the single-agent MAB problem was solved in a scenario involving the integration of multiple radio access technologies into edge computing. Wu \cite{ref11} extended the MAB framework to address the multi-agent MAB problem in both centralized and distributed bandit settings. 
Furthermore, the application of contextual-combinatorial MAB models, an extension of the basic MAB, was employed to address failures or delays in edge computing \cite{ref23}. 
These studies demonstrate the versatility and effectiveness of MAB-based approaches in optimizing edge computing systems.


In this paper, we investigate an edge computing network operating within a 5G scenario, where multiple end users are interconnected with multiple ESs. To address the optimization problem, we employ the widely adopted multi-armed bandit (MAB) framework. We propose an adaptive task offloading algorithm (ATOA), specifically tailored for networks characterized by tidal data traffic patterns. We evaluate the performance of the proposed algorithms and find that ATOA consistently outperforms both the $\varepsilon$-greedy algorithm and the UCB1-based algorithm in the same simulation scenario.


The remainder of this paper is organized as follows. Section \ref{system model and problem formulation} presents the system model and formulates the MAB problem. In Section \ref{algorithms}, we utilize the $\varepsilon$-greedy algorithm and the UCB1-based algorithm to tackle the formulated problem. Section IV presents the details of our proposed ATOA algorithm. The simulation results and a comparative analysis of the performance of the proposed algorithms are provided in Section \ref{results}. Finally, Section \ref{conclusion} concludes the paper.

\section{System Model And Problem Formulation}
\label{system model and problem formulation}

In this section, we present the system model and formulate the resource allocation problem using the MAB approach.

\subsection{System Model}

\begin{figure}[h] 
\setlength{\abovecaptionskip}{0.cm}
\setlength{\belowcaptionskip}{-0.cm}
\centering
\includegraphics[width=0.8\columnwidth]{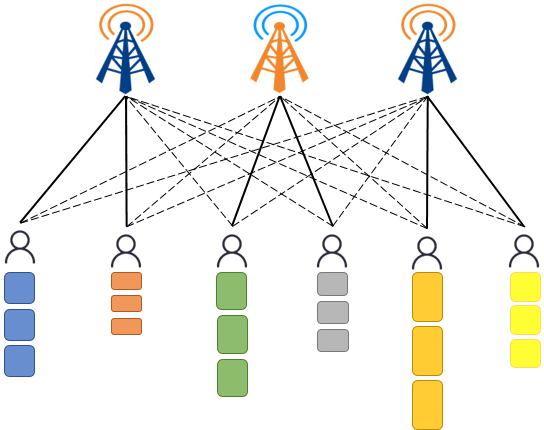}
\caption{A multi-user multi-server edge network.} 
\label{fig 1: system model}
\end{figure}

As shown in Fig. \ref{fig 1: system model}, we consider a two-layer edge computing network where a set of $I$ users $\mathcal{I}=\{1,2,\ldots,I\}$ are connected to a set of $J$ ESs $\mathcal{J}=\{1,2,\ldots,J\}$. Each ES is a Radio Access Network (RAN) node, either a gigabit NodeB (gNB, i.e., a 5G base station) or an enhanced NodeB (eNB, i.e., a 4G base station). Due to the fact that the number of end devices is much larger than the number of base stations in a real 5G scenario, here we have $I > J$. Moreover, because of the variety of smart devices, these end users have different and limited computation capacities, denoted by $\mathcal{C}^u=\{C_1^u,C_2^u,\ldots,C_I^u\}$. Similarly, the ESs have computation capacities denoted by $\mathcal{C}^s=\{C_1^s,C_2^s,\ldots,C_J^s\}$.


Consider a time-slotted system with $T$ slots, indexed by the set $\mathcal{T}=\{1,2,\ldots,T\}$. At each time slot $t \in \mathcal{T}$, each user $i$ generates a computation task $\mathcal{L}_{i,t}$ of size $S_{i,t}$. All the generated tasks $\{\mathcal{L}_{i,t}\}_{i\in \mathcal{I}}$ are delay-sensitive and need to be performed within a certain execution deadline of $\tau$. Therefore, the size $\hat{S}_i$ of the task that user $i$ can process locally within the duration $\tau$ is given as
\begin{equation}
    \hat{S}_i=\tau C_i^u/d,
\end{equation}
where $d$ represents the number of CPU cycles required to process 1 bit of data.

To enhance network resource utilization and reduce task processing latency, we propose a task processing mechanism as follows. For each user $i$, if the task $\mathcal{L}_{i,t}$ has a size $S_{i,t}\leq\hat{S}_i$ at time slot $t$, it is processed purely through local computing. Otherwise, we assume that the task $\mathcal{L}_{i,t}$ can be split into two parts. The first part $\hat{\mathcal{L}}_{i,t}$ with size $\hat{S}_i$ is processed locally by user $i$; the second part $\bar{\mathcal{L}}_{i,t}$ is offloaded to an ES $j\in\mathcal{J}$, with the size given as
\begin{equation}
    \bar{S}_{i,t} = S_{i,t}-\hat{S}_{i,t} = S_{i,t}-\tau C_i^u/d.
\end{equation}


Hence, each task $\mathcal{L}_{i,t}$ is executed in either of the following two ways, denoted by the flag $k_{i,t}\in\{-1,1\}$. To elaborate, we denote the processing latency of task $\mathcal{L}_{i,t}$ by $L_{i,t}$.

1) Execute $\mathcal{L}_{i,t}$ locally with $k_{i,t}=-1$: Given the computation capacity $C_i^u$, the task processing latency $L_{i,t}$, equal to the computing latency, is hence given as
\begin{equation}\label{local latency}
    L_{i,t} = d S_{i,t}/C_i^u.
\end{equation}

2) Execute $\hat{\mathcal{S}}_i$ locally and offload $\bar{\mathcal{L}}_{i,t}$ to ES $j$ with $k_{i,t}=1$: In this case, the total task processing latency consists of the computation latency and the transmission delay from user $i$ to ES $j$. As a result, it can be calculated as
\begin{eqnarray}\label{server latency}
    L_{i,t}=\tau+\bar{S}_{i,t}/R_{i,j} + d\bar{S}_{i,t}/C_j^s,
\end{eqnarray}
where $R_{i,j}$ is the data capacity between user $i$ and ES $j$.

With the focus on the completion of the tasks $\{\mathcal{L}_{i,t}\}_{i=1}^{I}$, we define the maximum latency $L(t)$ of all the tasks as a measure of the performance, given as  
\begin{equation}\label{max latency}
    L(t) = \max_{i\in \mathcal{I}} \{L_{i,t}\}.
\end{equation}

Our objective is to find an optimal task offloading policy that is able to minimize the task processing latency.

\subsection{Problem Formulation}
To start, we introduce the so-called MAB framework. Consider a situation where a gambler faces a slot machine with multiple arms, and the rewards of all the arms are unknown beforehand. Once an arm of the slot machine has been played at a time slot, the gambler can obtain a reward. The goal of MAB is to learn the optimal arm from a  set of candidate arms by selecting and observing obtained rewards. The gambler endeavors to maximize the long-term reward within limited time slots\cite{ref23}.

This work adopts the MAB framework to solve the task offloading problem. In the edge network under study, the users are treated as agents to learn the global optimal task offloading strategy by making a choice between exploration (i.e., pulling an arm that has never been chosen) and exploitation (i.e., pulling an arm that has been proven the best empirically).

Particularly, at time slot $t$, the system pulls an arm $\mathbf{A}_n=\left\{a_{1,n},a_{2,n},\ldots,a_{I,n}\right\}$ from the set $\mathcal{A}=\{\mathbf{A_1},\mathbf{A_2},\ldots,\mathbf{A_N}\}$ which includes all the possible task-processing choices that the users can make at any given time slot. The arm $a_{i,n}$ is defined as
\begin{equation}
    a_{i,n} = \begin{cases}
     -1 & \text{if $k_i^t=-1$},\\ j,& \text{if $k_i^t=1$.}
     \end{cases}
\end{equation}
where $a_{i,n}$ is set to $j$ if user $i$ selects ES $j$ for task offloading with $k_i^t=1$, and $a_{i,n}$ is set to -1 with local computing.





We define the reward function $R(t)$ as the task processing latency $L(t)$, which represents a negative reward used to evaluate the task offloading decisions made by all end users. Our objective is to minimize the long-term reward, ultimately aiming to minimize the cumulative task processing latency over the given $T$ time slots. It can be formulated as
\begin{equation}
    \mathop{\min} \sum_{t=1}^T \gamma^{T-t} R_{\mathbf{A_n}}(t),
\end{equation}
where $R_{\mathbf{A_n}}(t)$ is the observed reward of the selected arm $\mathbf{A_n}$ at time slot $t$ and $\gamma$ is the discount factor.

\section{Conventional MAB Algorithms for Task Offloading}
\label{algorithms}
In this section, two conventional MAB algorithms, namely $\varepsilon$-greedy algorithm and UCB1-based algorithm, are adopted to address our formulated task offloading problem. Then we test and compare their performance in different scenarios.

\subsection{$\varepsilon$-greedy Task Offloading Algorithm}
In the $\varepsilon$-greedy method, an agent faces a decision-making choice and can opt for exploration with a probability of $\varepsilon \in [0,1]$, or exploitation with a probability of 1-$\varepsilon$. Exploration involves randomly selecting an arm from the set of unexplored options, while exploitation entails pulling the optimal arm with the best-known value among the explored options. 
Thus, the value of $\varepsilon$ plays a crucial role in balancing exploration and exploitation. 
A larger $\varepsilon$ increases the likelihood of exploration, while a smaller value indicates a preference for exploitation.

Based on the $\varepsilon$-greedy method, we propose the $\varepsilon$-greedy task offloading algorithm to address the task offloading problem in a multi-user multi-server edge computing network. The first $S$ time slots serve as the exploration phase. At each time slot $t$ ($t\leq S$), one of the unexplored arms is randomly selected and then its reward is calculated using Eq.(\ref{local latency}, \ref{server latency}, \ref{max latency}).

In the subsequent ($T-S$) time slots, both exploration and exploitation take place. Specifically, there is a probability of $\varepsilon$ to randomly choose an arm from the set of unexplored options, while there is a probability of $1-\varepsilon$ to select the optimal arm with the minimum negative reward according to the observed information during the exploration stage. 


\subsection{UCB1-Based Task Offloading Algorithm}

Different from the $\varepsilon$-greedy algorithm, the classic UCB1 method considers not only the reward of each arm but also the number of times an arm has been selected in previous time slots. Therefore, based on the UCB1 arm selection formula \cite{ref16}, we determine the optimal arm at time slot $t$ using the estimate reward $\hat{R}_{\mathbf{A_n}}(t)$, defined as
\begin{equation} \label{expec_reward}
    \hat{R}_{\mathbf{A_n}}(t)=\overline{R}_{\mathbf{A_n}}(t-1)-U\sqrt{\frac{\xi\ln{t}}{1+N_{\mathbf{A_n}}(t)}}.
\end{equation}
Here $\hat{R}_{\mathbf{A_n}}(t)$ denotes the estimate reward that arm $\mathbf{A_n}$ brings along at time slot $t$, while $\bar{R}_{\mathbf{A_n}}(t-1)$ represents the cumulative average reward of arm $\mathbf{A_n}$ in the previous ($t-1$) time slots. $U$ denotes the utmost amplitude of $R_{\mathbf{A_n}}(t)$, given as 
\begin{equation}
    U = \sup_{\forall t,\mathbf{A_n}} R_{\mathbf{A_n}}(t) - \inf_{\forall t,\mathbf{A_n}} R_{\mathbf{A_n}}(t).
\end{equation}
$U$ and $\xi$ are parameters designed to evaluate the confidence radius\cite{ref11}. Specifically, larger values of $U$ and $\xi$ increase the probability of exploration.
Furthermore, $N_{\mathbf{A_n}}(t)$ is the number of times that arm $\mathbf{A_n}$ has been selected before time slot $t$.


More precisely, the first $S$ time slots are dedicated to exploration. When an arm $\mathbf{A_n}$ is selected at time slot $t$, its current reward $R_{\mathbf{A_n}(t)}$ is calculated immediately, and $N_{\mathbf{A_n}(t)}$ as well as its cumulative average reward $\overline{R}_{\mathbf{A_n}}(t-1)$ are updated accordingly. During the remaining ($T-S$) time slots for exploitation, the expected rewards of all arms in $\mathcal{A}$ are estimated initially at each time slot $t$. Since the reward is negative in the proposed framework, the arm with minimum estimated reward is regarded as optimal. 
Once the optimal arm is selected, the policy observes its total task processing latency and updates its average reward $\overline{R}_{\mathbf{A_{opt}}}(t)$ afterward.

\begin{figure}[t!]
    \centering
    \includegraphics[width=0.4\textwidth]{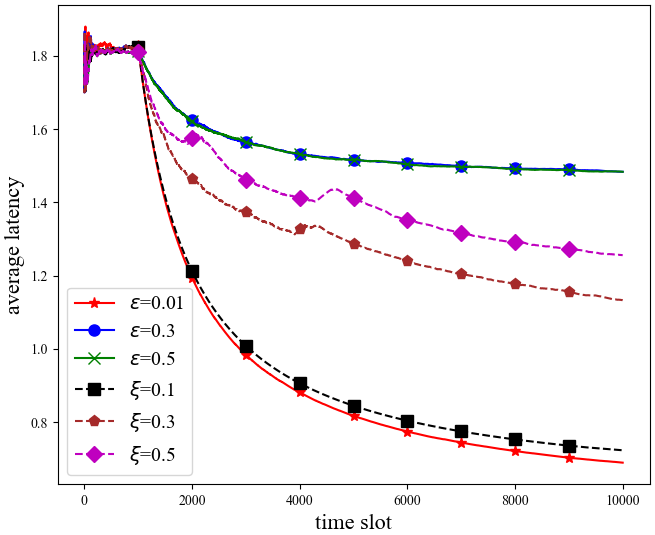}
    \caption{Performances of the $\varepsilon$-greedy Algorithm and the UCB1-Based Algorithm in the stable scenario}
    \label{fig 5: stable}
\end{figure}

\begin{figure}
    \centering
    \includegraphics[width=0.4\textwidth]{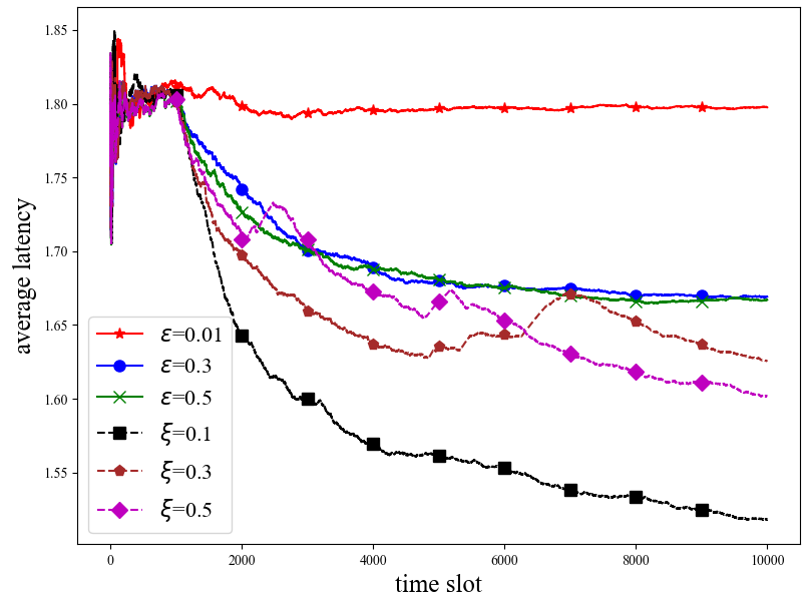}
    \caption{Performances of the $\varepsilon$-greedy Algorithm and the UCB1-Based Algorithm in the unstable scenario}
    \label{fig 6: unstable}
\end{figure}

\subsection{Performance of the Conventional MAB Algorithms}
To assess the performance of the two algorithms, we conduct experiments on edge computing networks under both stable and unstable scenarios.


In the stable network scenario, where the task size $S_{i,t}$ ($t\in \mathcal{T}$) remains relatively constant, we compare the performance of the two algorithms with different values of $\varepsilon$ and $\xi$, respectively, as shown in Fig.\ref{fig 5: stable}. It is obvious that as the value of $\varepsilon$ decreases, the performance of the $\varepsilon$-greedy algorithm improves, and the UCB1-based algorithm also demonstrates enhanced performance with decreasing $\xi$. This suggests that once sufficient environmental information is obtained, prioritizing exploitation over exploration yields better results. Moreover, both algorithms can exhibit can achieve similar levels of task-offloading efficiency.


In the case of an unstable network with erratic data traffic, although there is some performance degradation, the UCB1-based algorithm still outperforms other algorithms with $\xi=0.1$, as observed in Fig.\ref{fig 6: unstable}. However, the $\varepsilon$-greedy algorithm with $\varepsilon=0.01$ struggles to reduce the average task processing latency in this scenario, indicating its poor robustness compared to the UCB1-based algorithm.


\textbf{Remark}: We can infer that the $\varepsilon$-greedy method is more suitable for stable networks, whereas the UCB1-based method is better suited for networks with unstable data traffic. Furthermore, the $\varepsilon$-greedy method is relatively easy to implement, while the UCB1-based method introduces higher computational complexity. These observations inspire the development of an adaptive method that leverages the benefits of both approaches.


\section{ Adaptive Task Offloading Algorithm}\label{ATOA}

We proceed to introduce the proposed adaptive task offloading algorithm (ATOA), allowing for more flexible and effective task offloading in various network scenarios. 

To elaborate, at time slot $t$, we first calculate the average variance $v_t$ of task sizes over the last $D$ time slots, given as
\begin{equation}
    v_t = \frac{1}{I} \sum\limits_{i=1}^I\left[ {\frac{1}{D}\sum\limits_{q = t-D}^{t-1} {\left( {S_{i,q} - m_i} \right)^2 }}\right],
\end{equation}
where $m_i$ is the average size of tasks generated by user $i$ over the last $D$ time slots. 

Based on the value of variance $v_t$, ATOA predicts whether the network traffic state $s_t$ of the current time slot is stable, defined as 
\begin{equation}
  s_t = \begin{cases}
 -1, & \text{if $v_t>a$}, \\ 1, & \text{if $v_t\leq a$.}
 \end{cases}
\end{equation}
Here $a$ is a threshold used to evaluate the stability of the network data traffic. When the average variance $v_t\leq a$, ATOA considers the network state as stable ($s_t=1$). Otherwise, the current network state is classified as unstable ($s_t=-1$). 



In the exploration phase, in addition to selecting a random arm and calculating its immediate reward, the memory pool is updated with the sizes of newly generated tasks. At each time slot when exploiting, the first step is to predict the current network traffic state according to those task sizes stored in the memory pool. If the prediction turns out to be stable, $\varepsilon$-greedy method would be adopted for task offloading. Otherwise, the task offloading strategy is developed using UCB1-based method. The detailed process of ATOA is outlined in \textbf{Algorithm\ref{Algorithm 3}}.

\begin{algorithm}[t!]
   \caption{Adaptive Task Offloading Algorithm\label{Algorithm 3}}
   \begin{algorithmic}[1]
    \State Initialization: $\varepsilon,\xi;\mathcal{A}$=$\left\{\mathbf{A_1},...,\mathbf{A_N}\right\};N_{\mathbf{A_n}}=0.$
    \State \textbf{Exploration}:
      \For {$t=1,\ldots,S$}
      \State Choose an arm $\mathbf{A_n}$ from arm space $\mathcal{A}$;
     \State Calculate $R_{\mathbf{A_t}}(t)$;
     \State Update memory pool with current network traffic state;
     \EndFor
     \State \textbf{Exploration-Exploitation}:
     \For {$t=S+1,\ldots,T$}
     \State Predict network state $s_t$ according to variance $v_t$;
     \If{$s_t=1$}
     \State Adopt $\varepsilon$-greedy method:
     \State Generate a random number $randnum$;
     \If {$randnum < \varepsilon$}
     \State Choose an arm $\mathbf{A_n}$ from $\mathcal{A\textbackslash A'}$ randomly;
     \State Add the chosen arm $\mathbf{A_n}$ into $\mathcal{A'}$;
     \State Calculate $R_{\mathbf{A_n}}(t)$;
     \Else
     \State Choose optimal arm $\mathbf{A_{opt}}=\mathop{\arg\min}\limits_{\mathbf{A_n}\in\mathcal{A'}} R_{\mathbf{A_n}}(t)$;
     \State Calculate $R_{\mathbf{A_{opt}}}(t)$;
     \EndIf     
     \Else
     \State Adopt UCB1-based method:
     \For {$n=1,\ldots,N$}
     \State Estimate expected reward $\hat{R}_{\mathbf{A_n}}(t)$ in (\ref{expec_reward});
     \EndFor
     \State Choose the optimal arm $\mathbf{A_{opt}}=\mathop{\arg\min}\limits_{\mathbf{A_n}\in\mathcal{A}}\hat{R}_{\mathbf{A_n}}(t)$;
     \State Calculate $R_{\mathbf{A_{opt}}}(t)$; 
     \State $N_{\mathbf{A_n}}=N_{\mathbf{A_n}}+1$;
     \EndIf
     \State Update cumulative average reward and memory pool;
     \EndFor
    \end{algorithmic}
\end{algorithm}


\textbf{\begin{figure*}[t!]
    \centering
    \begin{minipage}[b]{0.34\textwidth}
        \centering
        \includegraphics[width=1\textwidth]{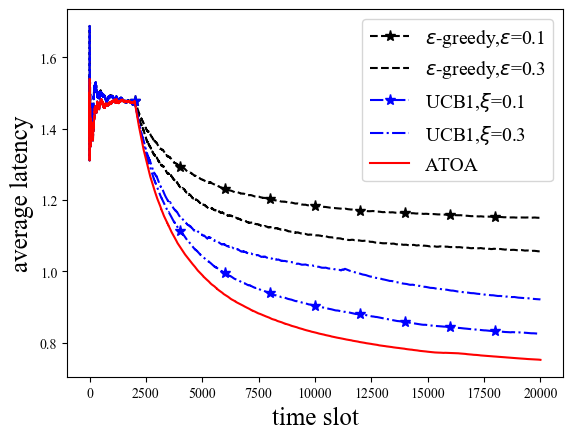}
    \caption{Performance comparison: the $\varepsilon$-Greedy, UCB1-Based Algorithm, and ATOA}
    \label{tidal_3alg}
    \end{minipage}
    \begin{minipage}[b]{0.32\textwidth}
        \centering
        \includegraphics[width=1\textwidth]{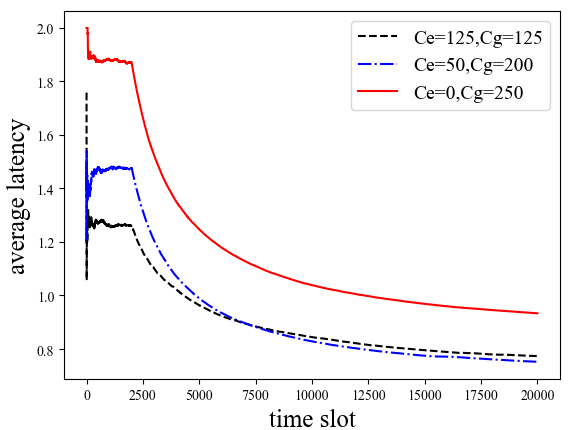}
    \caption{Performance of ATOA with different values of $C_e$ and $C_g$}
    \label{capacity_vary}
    \end{minipage}
    \begin{minipage}[b]{0.32\textwidth}
        \centering
        \includegraphics[width=1\textwidth]{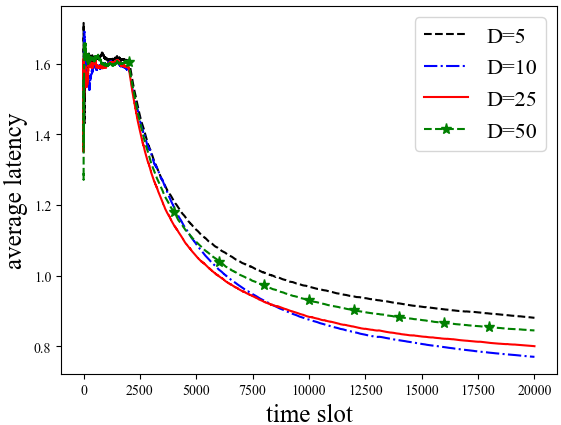}
    \caption{Performance of ATOA with different values of $D$}
    \label{D_vary}
    \end{minipage}
\end{figure*}}

\section{Simulation Results}\label{results}

In this section, we present simulation experiments and performance comparisons.

We consider a scenario characterized by tidal data traffic patterns. Specifically, the data traffic state exhibits stability for a period of $T_s$ time slots, during which each ES processes tasks with a relatively constant size. We assume that the overall task size $S_{i,t}$ across the time slots follows a normal distribution with mean $\mu_i=10i$ and variance $\sigma_{i,1}=\mu_i/10$. On the contrary, the state of data traffic is more dynamic and unstable in the following $T_u$ time slots, for which we set the parameters of task size as $\mu_i$ and $\sigma_{i,2}=\mu_i/2$.
Consider the edge computing network with $I=6$ end users and $J=2$ ESs. ES $1$ is set as a gNB with computation capacity $C_1^s=C_g$ while ES $2$ is an eNB with $C_2^s=C_e$. 
Moreover, the operations are carried out over $T=20000$ time slots, with the first $S=10000$ time slots for exploration and the remaining ones for exploitation. Here we set $T_s$ and $T_u$ to both 150.

In addition, as $v_t$ represents the average variance of task sizes over the last $D$ time slots, the stability threshold $a$ should be determined based on both $D$ and the mean of task sizes $\mu_i$. Therefore, we define $a$ as $a=0.5D \sum{\mu_i} / I, i\in\mathcal{I}$ when implementing ATOA. 

With $D=10$, $C_e=50$, and $C_g=200$, we evaluate the performance of the $\varepsilon$-greedy algorithm, the UCB1-based algorithm, and the proposed ATOA algorithm, as depicted in Fig.\ref{tidal_3alg}. In ATOA, we set $\varepsilon'$ and $\xi'$ to 0.01 and 0.1, respectively. This is based on the observation that the $\varepsilon$-greedy algorithm performs better with a small $\varepsilon$ in a generally stable network (Fig.~\ref{fig 5: stable}), and the UCB1-based algorithm is more robust with a small $\xi$ in an unstable network (Fig.~\ref{fig 6: unstable}).
It is evident that ATOA outperforms both of them. Once in the exploitation phase, ATOA achieves the fastest decrease in average latency and maintains the lowest latency at time slot 20000. Additionally, since ATOA adopts the $\varepsilon$-greedy method during periods of stable data traffic, the computational complexity is also reduced compared to the UCB1-based algorithm. These findings highlight the superiority of our proposed ATOA algorithm.

Fig.\ref{capacity_vary} shows a comparison of ATOT with varying differences between $C_e$ and $C_g$. We observe that as the difference between $C_e$ and $C_g$ increases, there is an even greater decrease in average task processing latency. This can be attributed to the fact that with a larger difference, the gap between the latency returned by "bad arms" and that returned by the optimal arm also widens, allowing ATOA to make more informed and advantageous decisions. However, regardless of the difference in values between $C_e$ and $C_g$, our proposed ATOA can effectively reduce the task processing latency to convergence.


Finally, Fig.~\ref{D_vary} illustrates the performance of ATOA with different values of $D$. It is shown that ATOA performs the best when $D$ is set to $10$, followed by $25$, and then $50$ and $5$. When $D$ is set to $5$, it is too small to capture sufficient information about the traffic state of the network. This limitation may result in incorrect predictions of the network state $s_t$ and subsequently lead to the selection of an inappropriate arm. Conversely, a value of $D=50$ is too large, which can also yield suboptimal results. Therefore, it is crucial to choose an appropriate value of $D$ when implementing our proposed ATOA algorithm.

\section{Conclusion} \label{conclusion}
In this paper, we have applied the MAB framework to optimize the task offloading strategies in edge computing networks with limited resources and varying conditions. To reduce task processing latency, we have developed the $\varepsilon$-greedy algorithm and the UCB1-based algorithm. The simulation results have demonstrated that the former is well-suited for stable networks and offers ease of implementation, while the latter is of stronger robustness with high levels of instability. By leveraging the strengths of both algorithms, we have proposed an adaptive algorithm called ATOA that achieves superior performance in terms of task offloading efficiency. This research contributes to the field of edge computing by providing practical solutions for optimizing task offloading strategies in diverse network conditions.


\bibliographystyle{IEEEtran}
\bibliography{reference.bib}
\end{document}